# Electronic and optical properties of quantum wells embedded in wrinkled nanomembranes

P. Cendula,[1,2] S. Kiravittaya,[1] O. G. Schmidt[1,3]

[1]Institute for Integrative Nanosciences, IFW Dresden, Helmholtzstrasse 20, 01069 Dresden, Germany
[2]DCMP, Faculty of Mathematics and Physics, Charles University, Ke Karlovu 5, 12116 Prague, Czech Republic
[3]Material Systems for Nanoelectronics, Chemnitz University of Technology, Reichenhainer Strasse 70, 09107 Chemnitz, Germany

The authors theoretically investigate quantum confinement and transition energies in quantum wells (QWs) asymmetrically positioned in wrinkled nanomembranes. Calculations reveal that the wrinkle profile induces both blue- and redshifts depending on the lateral position of the QW probed. Relevant radiative transistions include the ground state of the electron (hole) and excited states of the hole (electron). Energy shifts as well as stretchability of the structure are studied as a function of wrinkle amplitude and period. Large tunable bandwidths of up to 70 nm are predicted for highly asymmetric wrinkled QWs.

Wrinkled (or buckled) nanomembranes[1] have recently been investigated to provide stretchability/bendability for a broad range of applications such as integrated circuits,[2] magnetoelectronics,[3] photodetectors,[4] solar cells[5] and light emitting diodes.[6] Incorporating quantum wells (QWs) into a flexible thin membrane, for instance, enables strain to modify the magnetoresistive[7] or optical properties[8] of the structure. In particular, large energy shifts caused by strain modulations might be interesting for broadband QW detectors and emitters.[9] However, there is little work elucidating the limits and impact of modulated strain on the emission properties of wrinkled nanomembranes incorporating QWs.

In this work, we study the electronic and optical properties of wrinkled nanomembranes incorporating QWs at different positions away from the mechanical neutral plane of the nanomembrane. The wrinkle shape provides strain in both concave and convex regions, which alters the bandedge potential and causes a pronounced lateral variation of the transition energies. Large wavelength tunability is predicted and scaling properties of the structure are discussed.

As a typical example, we consider a nanomembrane of total thickness $d$ consisting of an $In_{0.22}Ga_{0.78}As$ QW layer ($4\,\text{nm}$) with initial misfit strain $\varepsilon_0 = -1.5\%$ sandwiched between two GaAs barrier layers with variable thicknesess and $\varepsilon_0 = 0$. A total thickness of $d = 44\,\text{nm}$ is assumed and the GaAs barrier thickness is always larger than 10 nm in order to ensure good carrier confinement in the QW layer.[10] The nanomembrane can be fully[2] or partially[4] bonded to a prestrained flexible substrate and upon release of the prestrain, it deforms into wrinkles, which can be described by $A(1-\cos(kx))/2$, where $A$ and $k$ are wrinkle amplitude and wavenumber ($k = 2\pi/L$, $L$ is wrinkle period), respectively, Fig. 1(a). We consider $A = 50\,\text{nm}$ and $L = 1\,\mu\text{m}$ as realistic and typical parameters[4] and periods $L$ down to $\approx 300\,\text{nm}$ have been reported, previously.[11]

The lateral strain $\varepsilon_x$ in the wrinkled QW (WQW) consists of the initial misfit strain and the bending strain i.e. $\varepsilon_x = \varepsilon_0 - zAk^2\cos(kx)/2$, where $z$ is the coordinate along the membrane thickness measured from the neutral plane of the membrane,[12] Fig. 1(a). The QW is placed at a distance $r$ away from the neutral plane and the QW is called symmetric for $r = 0$ and asymmetric for $r > 0$. The maximum bending strain in the convex and concave QW region can be written as $\varepsilon_{QW} = \pm 2\pi^2 r'A'/L'^2$ and it depends only on scaled variables $r' = r/d$, $A' = A/d$, and $L' = L/d$. The membrane fractures when the surface bending strain $\varepsilon_{surf}$ reaches the fracture limit $\varepsilon_{frac}$ (e.g. $\approx 2\%$ for GaAs[4]), $\varepsilon_{surf} = \pi^2 A'/L'^2 = \varepsilon_{frac}$, implying that the maximum amplitude before fracture is $A'_{max} = \varepsilon_{frac}L'^2/\pi^2$. The maximum bending strain of 2% (even if its deformation is large) implies that the curvilinear coordinate grid of the membrane is only slightly deformed and we can approximate it with the cartesian system $(x,y,z)$, Fig. 1(a). Because the membrane is thin, strain along the thicknesess $\varepsilon_z$ is given by $-\nu(\varepsilon_x + \varepsilon_y)/(1-\nu)$ [12] and the volumetric strain is $\varepsilon_{vol} = (1-2\nu)(\varepsilon_x + \varepsilon_y)/(1-\nu)$, where $\nu$ is Poisson's ratio. We keep the strain in $y$ direction constant, $\varepsilon_y = \varepsilon_0$.

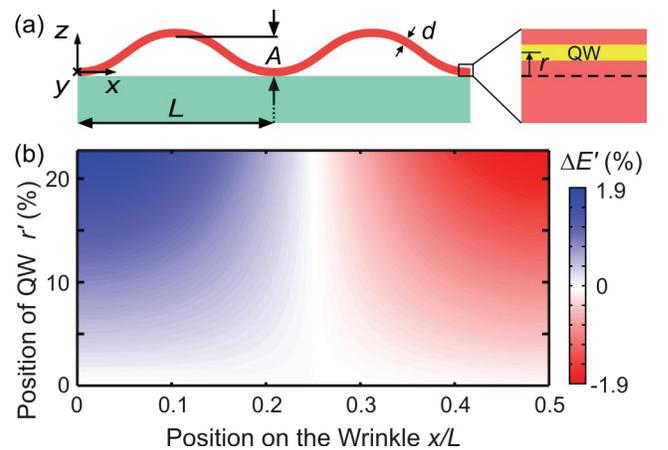

Figure 1: (a) Sketch of the WQW fixed in valleys to a stretchable substrate (not to scale). Inset shows the layer structure, neutral plane (dashed) and distance of QW from neutral plane $r$. (b) Transition energy shift $\Delta E'$ as a function of $x/L$ (only half period is shown) and $r'$, for typical parameters mentioned in text.

The lateral strain $\varepsilon_x$ linearly changes the bandedge energy,[13] with opposite signs in convex and concave WQW regions. The shift of the conduction and valence band with respect to the unstrained material is given by $a_C \varepsilon_{vol}$ and $a_V \varepsilon_{vol} - b_V \left( \varepsilon_z - \left( \varepsilon_x/2 + \varepsilon_y/2 \right) \right)$, where $a_C$, $a_V$, $b_V$ are the corresponding deformation potentials. Since the QW is compressively strained, we consider only heavy holes.[14]

The Schrödinger equation (SE) is numerically solved to obtain quantized energies of the electron and hole. An optical transition within the QW is due to creation/annihilation of an electron-hole pair with quantum numbers $n, m$, and photons with energy $E$ are absorbed/emitted. The transition rate is proportional to the square of the overlap integral $|O_{nm}|^2$ between electron and hole wavefunctions.[15] The shift in the transition energy relative to the energy of a planar strained QW $E_0$ is defined as $\Delta E' = (E - E_0)/E_0$.

In the one-dimensional (1D) case, we solve the SE in $z$ direction at a given point in $x$ by finite-difference method.[16] For a symmetric QW, small bending strain creates only little lateral variation of the transition energy $\Delta E'$, Fig. 1(b). However, an increase of bending strain for asymmetric QWs causes red and blue shifts of the transition energy $\Delta E'$ for convex and concave regions, respectively, which both increase in magnitude for increasing $r'$, Fig. 1(b). We apply our 1D model to experimentally reported wrinkled nanomembranes,[8] and compute a $-7.6\,\text{meV}$ shift for the concave region, which agrees well with the measured value of $-7.8\,\text{meV}$ [8] without any fitting parameter.

We proceed by considering the lateral confinement in the 2D cross-section of the WQW and solve the SE numerically. Modulation of strain in lateral direction modifies the conduction bandedge, Fig. 2(a), and thus a small lateral potential well is created in addition to the vertical potential well. Periodic boundary conditions were used and the Dirichlet condition was applied on the top and bottom surfaces of the WQW.

For the symmetric WQW, the electron and hole wavefunctions are both localized laterally in the convex and concave regions because these are equivalent, Fig. 2(b). For asymmetric WQWs, the ground state of the electron $\psi_{e0}$ is localized in the concave region and the ground state of the hole $\psi_{h0}$ in the convex region, Figs. 2(c,d). This is because the potential energy difference between the convex and concave region is proportional to $a_C \varepsilon_{QW} (1-2\nu)/(1-\nu)$ for electron and $c_V \varepsilon_{QW}$ for hole, where $a_C < 0$ and $c_V = [a_V(1-2\nu) - b_V(1+\nu)/2]/(1-\nu) > 0$ for the QW layer. Consequently, the potential minima for electron and hole are located in different WQW regions. The lateral confinement energy of the electron is typically weak (0.3 meV), and it scales as $h\sqrt{a_C \varepsilon_{QW}/m^*}/L$ when a parabolic well is assumed ($h$ is Planck constant, $m^*$ is effective mass). The maximum lateral confinement energy of $\approx 6\,\text{meV}$ is obtained for minimum $L = 300\,\text{nm}$ [11] and maximum $\varepsilon_{QW} \approx \varepsilon_{frac} = 2\%$.

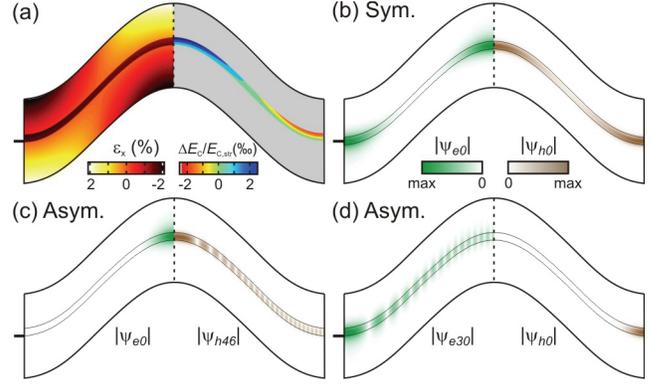

Figure 2: (a) Lateral strain ($\varepsilon_x$, left) and normalized shift of the conduction bandedge relative to the strained QW (right) for asymmetric WQW, $r' = 4.6\%$. Localization of electron (green) and hole (brown) wavefunctions for (b) symmetric WQW and (c,d) asymmetric WQW, $r' = 4.6\%$. Due to mirror symmetry, only half of the period is plotted. Positions of the neutral plane are shown as black ticks on left side of the membranes, whose thicknesses are enlarged for clarity.

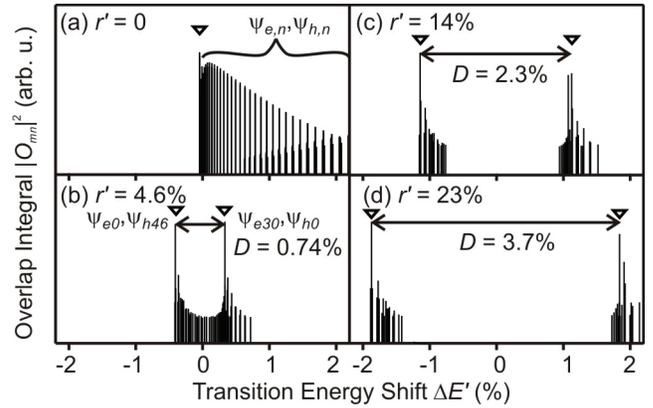

Figure 3: Bar plots of the overlap integrals versus $\Delta E'$ for increasing asymmetry $r'$ of the QW from (a) to (d). The transition bandwidth $D$ is labeled. Only transitions $n < 85$, $m < 120$ and $|O_{nm}|^2 > 0.1$ are shown.

Even though the ground state of the hole is not localized in the same region as the ground state of the electron, the excited hole states increasingly overlap with the electron ground state, Fig. 2(c). Excited electron states also overlap with the ground hole state, Fig. 2(d). The overlap integral for electron-hole transitions as a function of transition energy shift $\Delta E'$ is shown for increasing asymmetry $r'$ in Fig. 3. For $r' = 0$, the maximum overlap is mainly found for $n = m$ and transitions are blueshifted because excited states have higher energy. For $r' = 4.6\%$, the maximum overlap is between $\psi_{e0}$-$\psi_{h46}$ and $\psi_{e30}$-$\psi_{h0}$. Generally two groups of transitions are identified consisting of (1) high $\psi_{e,n}$ and first few $\psi_{h,m}$ in the convex region (blueshifted peaks) and (2) high $\psi_{h,m}$ and first few $\psi_{e,n}$ in the concave region (redshifted peaks), Figs. 3(b-d). The transition energy shift agrees well with our results from the 1D calculations ($\pm 1.9\%$ for $r'=23\%$), Fig. 1(b). Therefore, we can characterize the spectrum along the WQW by calculating the bandwidth $D$ between the minimum and maximum transition energy $\Delta E'$ along the WQW, Fig. 4(a). With increasing $L'$ or decreasing $A'$, the curvature effect becomes weaker and $D$ decreases. Fracture of the membrane (due to bending) will occur for $A' > A'_{max}$, at a strain of $\varepsilon_{surf} \approx 2\%$.[4] In general, the fracture strain $\varepsilon_{frac}$ depends on the strain rate and temperature[17] and can be as high as 7% for GaN.[18-19]

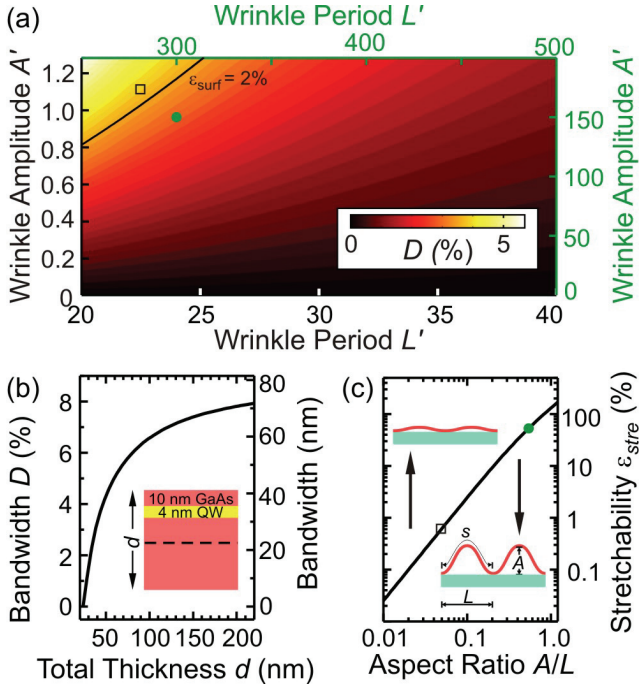

Figure 4: (a) Transition bandwidth $D$ as a function of normalized wrinkle period $L'$ and amplitude $A'$ for $r'=23\%$. Left and bottom axes are for typical WQW (square), top and right green axis are for scaled-up WQW (green circle), described in the main text. The condition of maximum $\varepsilon_{surf}$ of 2% is labeled. (b) Bandwidth $D$ (left axis) and wavelength (right axis) as a function of $d$ at the maximum $\varepsilon_{surf}=2.2\%$. (c) Stretchability $\varepsilon_{stre}$ of wrinkled membrane as a function of $A/L$ with square and circle for typical and scaled-up wrinkle parameters as in (a).

The maximum bandwidth $D_{max}$ can be estimated by recalling that $D$ scales linearly with strain $\varepsilon_{QW}$.[13] Up to here we have considered $d=44$ nm and $r'=23\%$, as compared to the maximum value $r'=50\%$ for large $d$. By increasing the total thickness, the bandwidth increases to $D \approx 8\%$ for $d=224$ nm, which corresponds to a variation in emission wavelenths of $\approx 70$ nm, Fig. 4(b). The upper limit of $D_{max} \approx 9\%$ is approached for very large $d$.

A highly stretchable WQW with large transition bandwidth is proposed, Fig 4(c). The stretchability $\varepsilon_{stre}$ is defined as $(s-L)/L$, where $s$ is the wrinkle arclength. For shallow wrinkles ($A \ll L$), the arclength can be approximated by $s \approx L\left(1+\pi^2 A^2/4L^2\right)$. Since the bandwidth $D$ scales with $\varepsilon_{QW}$ (which depends on $A$, $L$ and $d$) and the stretchability scales with aspect ratio $A/L$, they can be tuned independently. For example, if we scale up the wrinkle dimensions to $A'=150$, $L'=300$, we obtain two orders of magnitude larger stretchability, Fig. 4(c), and the same bandwidth, Fig. 4(a), as compared to typical wrinkle parameters used before. Because the bandwidth $D$ scales only with $\varepsilon_{QW}$, $D$ is the same for both scaled-up (green axes in Fig. 4(a)) and typical (black axes in Fig. 4(a)) wrinkle parameters, as they have the same $\varepsilon_{QW}$.

In conclusion, we have presented numerical studies of emission/absorption properties of wrinkled nanomembranes incorporating asymmetrically positioned single QWs. Calculations in 1D and 2D show consistent results, the latter providing lateral confinement and 2D wavefunction profiles. Transition energy shifts, tunable bandwidths and stretchability of the structure are discussed. Large transition energy shifts are obtained for highly asymmetric structures. Our results can be used for the design of bandwidth-tunable QW emitters or detectors and for different material systems.

We acknowledge valuable discussions with Y. F. Mei, V. M. Fomin, and C. Ortix. This work was supported by Volkswagen Foundation (I/84 072) and the U.S. Air Force (FA9550-09-1-0550).

Email: p.cendula@ifw-dresden.de